\documentclass[pre,amsmath,amssymb,showpacs]{revtex4}
\usepackage{epsf}
\usepackage{graphicx}
\usepackage{bm}
\begin{document}
\title{A paradigmatic model of Earth's magnetic field reversals}
\author{ F. Stefani, G. Gerbeth, U. G\"unther}
\affiliation{Forschungszentrum Rossendorf, P.O. Box 510119,
           D-01314 Dresden, Germany
}
% ------------------------------------------------------------------------
% ------------------------------ Abstract --------------------------------
\begin{abstract}
The irregular polarity reversals of the Earth's magnetic field
have attracted much interest during the last decades. 
Despite the fact that 
recent numerical simulations of the geodynamo 
have shown nice polarity transitions,
the very reason and the basic mechanism 
of reversals are far from being understood.
Using a paradigmatic mean-field dynamo model 
with a spherically symmetric
helical turbulence parameter $\alpha$ we attribute
the essential features of reversals
to the magnetic field dynamics
in the vicinity of an
{\it exceptional point} of the spectrum
of the non-selfadjoint dynamo operator. At such 
exceptional (branch) points of square
root type two real
eigenvalues
coalesce and continue as a complex conjugated pair of
eigenvalues. 
Special focus is laid on the comparison of 
numerically computed time series with paleomagnetic 
observations.
It is shown that the considered dynamo model with high 
supercriticality can explain the observed time scale and 
the asymmetric shape of 
reversals with
a slow decay and a fast field recovery. 
\end{abstract}
\maketitle
% --------------------------- Introduction --------------------------------

\section*{Introduction.}      %% use full stop ``.'' after section titles

There is ample paleomagnetic evidence that the 
axial dipole component of the Earth's magnetic 
field has reversed its polarity many times. The last
reversal, the Brunhes-Matuyama transition, occurred approximately
780000 years ago.
The mean rate of reversals varies from nearly
zero during the Kiaman and Cretaceous superchrons
to approximately 5 per Myr in the present.
Some observations suggest a pronounced {\it asymmetry} of 
reversals with the decay of the dipole of a given polarity being 
much slower than
the following recreation of the dipole with opposite
polarity \cite{VALET,VALET2005}.
Observational data also indicate a possible {\it correlation} of the
field intensity 
with the interval  between subsequent 
reversals \cite{VALET2005,TARDUNO}. 
A third
hypothesis concerns the {\it bimodal distribution}
of the Earth's virtual axial dipole moment (VADM) with two peaks
at approximately 4 $\times$ 10$^{22}$ Am$^2$
and at  twice that value \cite{PERRIN,SHCH,HELLER}.

Although these reversal  features are still
controversially discussed
in the literature, it is worthwhile to ask if and how
they could be represented within geodynamo theory.
With view on the recent dynamo experiments
in Riga and Karlsruhe \cite{RMP} one could also ask about 
the 
most essential ingredient for a dynamo experiment to 
exhibit irregular  reversals
in a similar way as the geodynamo does.

\begin{figure}[t]
\begin{center}
%   following commands fit figure into full page width..
\unitlength=\textwidth
\includegraphics[width=\textwidth]{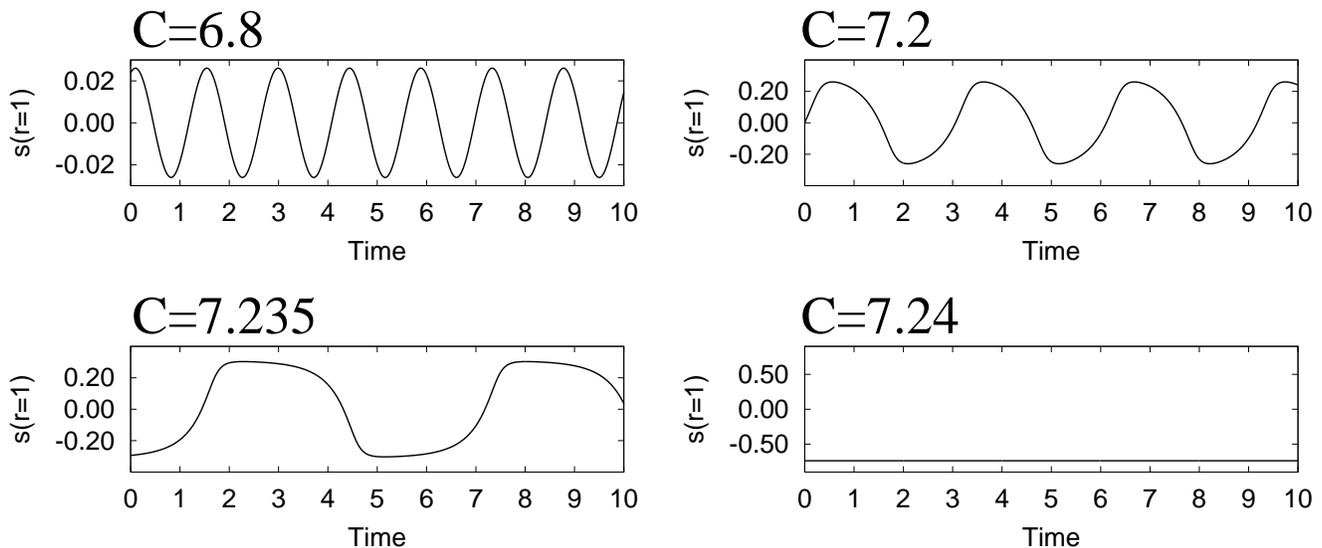}
\caption{Magnetic field evolution for $D=0$. With higher values 
of $C$, the field amplitude increases and the oscillation frequency 
decreases. Note also the growing anharmonicity 
(saw-tooth shape) of the oscillations and the transition 
to a non-oscillatory 
regime (as seen for $C=7.24$).}
\end{center}
\end{figure}

In a recent paper \cite{PRL} we had shown that  
a simple  mean-field dynamo model with a spherically symmetric
helical turbulence parameter $\alpha$
can exhibit all three mentioned features of reversals.
Interestingly, all of them  are  
attributable to the magnetic field
dynamics in the vicinity of an
{\it exceptional point} \cite{KATO} of the spectrum
of the non-selfadjoint dynamo operator where two real
eigenvalues
coalesce and continue as a pair of complex conjugate
eigenvalues. Typically, 
this exceptional point is associated with
a nearby local maximum of the
growth rate
dependence on the magnetic Reynolds number.
It is the negative slope of this curve between the local
maximum and the exceptional point that makes even 
stationary dynamos vulnerable to some prevailing noise.
This way, the system can leave the stable state and
run towards the exceptional point and beyond into the
oscillatory branch where the polarity transition occurs.

A weakness of this reversal model
was the apparent necessity to fine-tune the
magnetic Reynolds number and/or
the radial profile $\alpha(r)$
in order to adjust the operator spectrum in an appropriate way.
In a follow-up  paper \cite{EPSL} it was shown, however, that
this fine-tuning is not necessary in the case of {\it higher
supercriticality} of the dynamo.
It turned out that for increasing magnetic Reynolds number
there is a strong tendency for the exceptional point and the
associated local maximum to
move close  to the zero growth rate line were the
indicated reversal scenario can be actualized.
Although exemplified 
by the simple  spherically symmetric $\alpha^2$ dynamo model,
the main idea of this ''self-tuning'' mechanism
of saturated dynamos into a reversal-prone state
seems transferable to other dynamos.
Hence, reversing dynamos might be much more
typical than what could be expected from a purely
kinematic perspective.

In the present paper we go one step further and compare
paleomagnetic data of five reversals from 
the last two million years
with numerical time series resulting from our model.
It will be shown that it is again the 
strong supercriticality of 
the considered dynamo models that may explain the typical
time scales of the  observed asymmetric  reversals.

\section{The model.}

We consider a simple
mean-field dynamo model of $\alpha^2$ type
with a supposed 
spherically symmetric, isotropic helical turbulence parameter $\alpha$
\cite{KRRA}. The induction equation for the magnetic field $\bf B$ reads
\begin{eqnarray}
{\dot{ \bf{B}}} ={\bf \nabla}
\times (\alpha {\bf{B}}) +
(\mu_0 \sigma)^{-1} \Delta {\bf{B}} \; ,
\end{eqnarray}
with magnetic permeability
$\mu_0$ and
electrical conductivity $\sigma$.
For the Earth's  core we will assume the diffusion 
time   $\tau_{diff}=\mu_0 \sigma R^2$ 
to be $\sim 200$ kyr.

\begin{figure}[t]
\begin{center}
%   following commands fit figure into full page width..
\unitlength=\textwidth
\includegraphics[width=\textwidth]{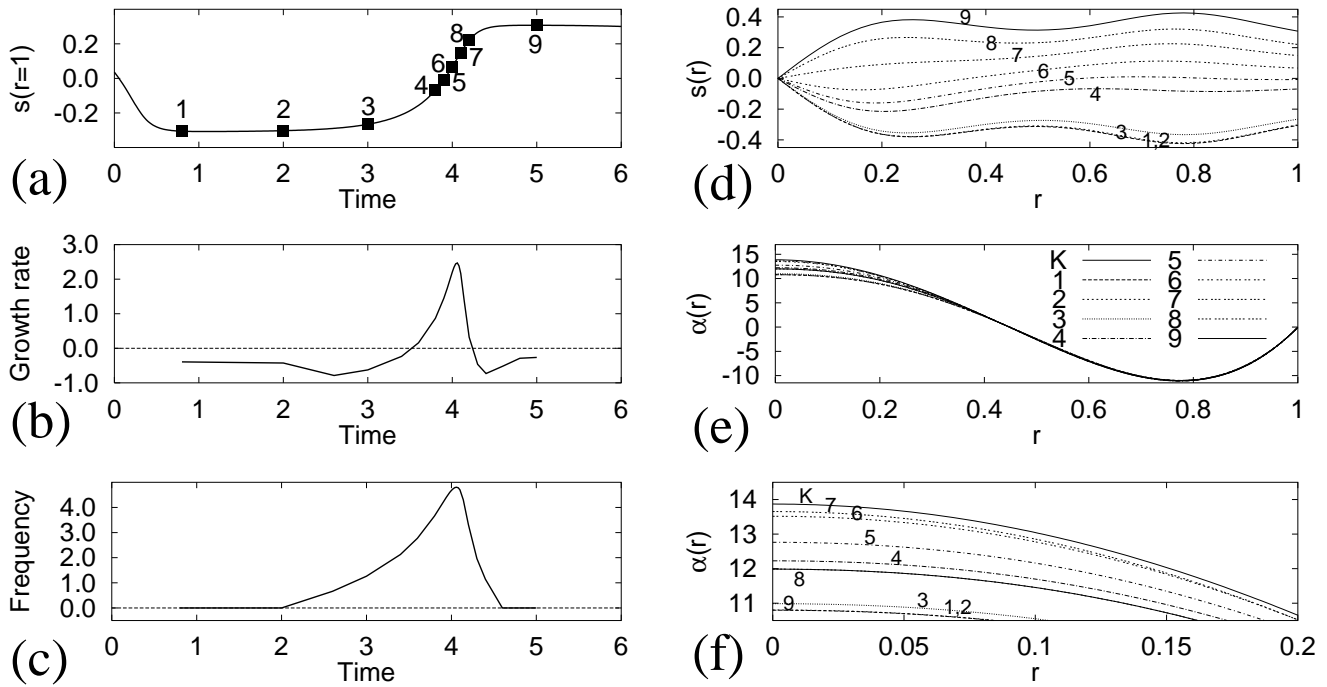}
\caption{Explanation of the field dynamics for $C=7.237$ and $D=0$. (a) 
Half an anharmonic oscillation with nine selected instants $\tau_i$, $i=1...9$. 
(b) 
Instantaneous growth rates resulting from the 
instantaneous $\alpha(r,\tau_i)$ profiles. (c) Instantaneous frequencies. 
(d) Instantaneous poloidal field $s(r,\tau_i)$. 
(e) Profiles $\alpha(r,\tau_i)$, 
compared to the unquenched $\alpha(r)$ (K). 
Note that the deformation  of $\alpha(r)$ during a reversal is not very
strong. (f) Details of (e) close to the origin. 
At the instants 6 and 7, the
$\alpha(r)$ profile comes close to the unquenched  $\alpha(r)$ (K).}
\end{center}
\end{figure}

The divergence-free magnetic field  ${\bf{B}}$ can be decomposed 
into a
poloidal and a toroidal components, according to
${\bf{B}}=-\nabla \times ({\bf{r}} \times
\nabla S)-{\bf{r}} \times
\nabla T $. The defining   scalars $S$ and $T$
are expanded in spherical harmonics of degree $l$ and order $m$
with expansion coefficients
$s_{l,m}(r,\tau)$ and $t_{l,m}(r,\tau)$.
For the envisioned spherically symmetric and isotropic $\alpha^2$ dynamo
problem,
the induction equation
decouples for each degree $l$ and order $m$ into the following pair
of equations:
\begin{eqnarray}
\frac{\partial s_l}{\partial \tau}&=&
\frac{1}{r}\frac{\partial^2}{\partial r^2}(r s_l)-\frac{l(l+1)}{r^2} s_l
+\alpha(r,\tau) t_l \; ,\\
\frac{\partial t_l}{\partial \tau}&=&
\frac{1}{r}\frac{\partial}{\partial r}\left( \frac{\partial}{\partial r}(r t_l)-\alpha(r,\tau)
\frac{\partial}{\partial r}(r s_l) \right)
-\frac{l(l+1)}{r^2}
[t_l-\alpha(r,\tau)
s_l] \; .
\end{eqnarray}
Since these equations are independent of the order $m$, we
have skipped $m$ in the index of $s$ and $t$.
The boundary conditions are
$\partial s_l/\partial r |_{r=1}+{(l+1)} s_l(1)=t_l(1)=0$.
In the following we consider only the dipole field with $l=1$.

\begin{figure}[t]
\begin{center}
%   following commands fit figure into full page width..
\unitlength=\textwidth
\includegraphics[width=\textwidth]{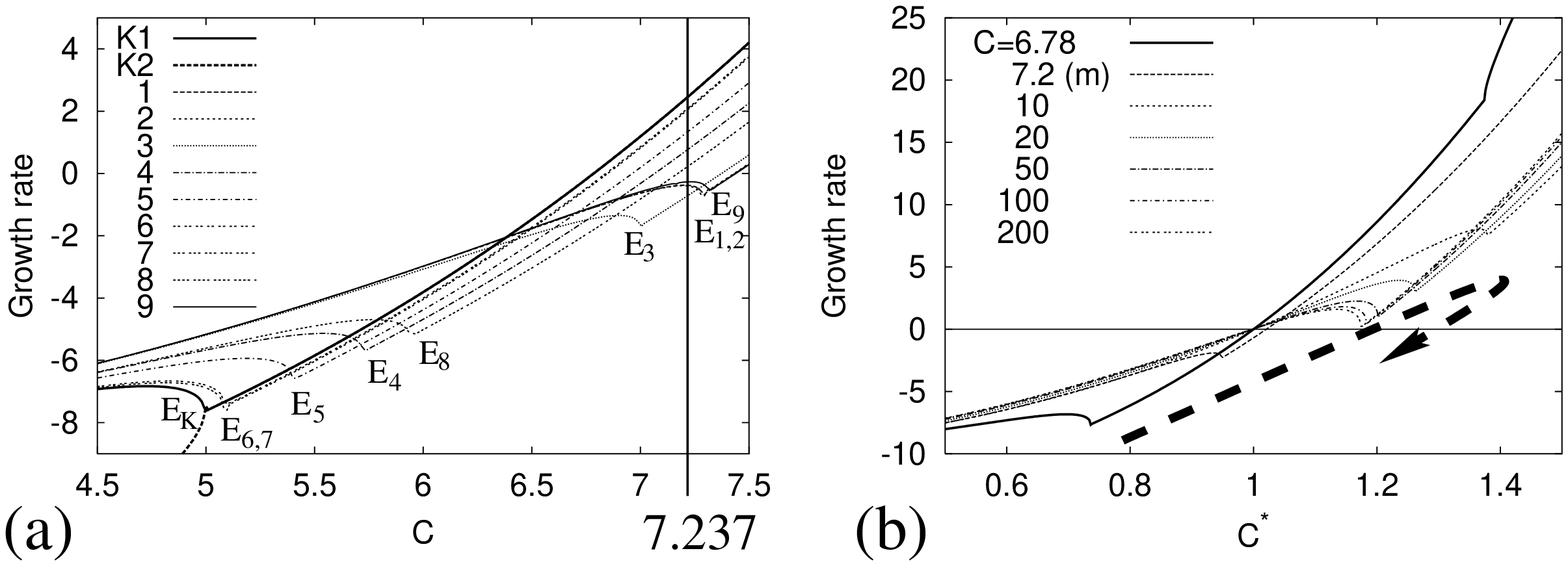}
 \caption{The role of exceptional points for the understanding 
of reversals. (a) Instantaneous growth rate curves for the kinematic profile
and for the quenched
$\alpha(r,\tau_i)$ profiles.
The $E_i$ indicate, for each of the considered
$\alpha$ profiles, the exceptional point. (b) Growth rates for the profiles 
$\alpha(r)= 1.916 \, C^* \, C  \, (1-6 \; r^2+5 \; r^4)/(1+E_{mag}(r)/E_{mag,0})$
with $C=$6.78, 7.2, 10, 20, 50, 100, 200.
The label ''7.2 (m)'' refers to
the maximally quenched $\alpha$ during the reversal. 
Note the remarkable move of the exceptional point
well above the zero line and back to it (indicated by the thick dashed line).}
\end{center}
\end{figure}

We will focus on 
a  particular radial profile 
$\alpha_{kin}(r)=1.916 \; C  \; (1-6 \; r^2+5 \; r^4)$,
which is characterized by a  sign change along the radius 
(the factor 1.916 results from normalizing the radial 
average of $|\alpha(r)|$ to the corresponding 
value for constant $\alpha$).
The motivation for this choice is that quite 
similar $\alpha(r)$ profiles 
had been shown to exhibit oscillatory behaviour \cite{OSZI}.

Saturation is ensured by assuming the kinematic
profile $\alpha_{kin}(r)$ to be algebraically 
quenched
by the  magnetic field energy averaged over the angles which
can be expressed in terms of $s_{1}(r,\tau)$ and $t_1(r,\tau)$.
Note that this averaging over the angles represents 
a severe simplification, 
since in reality 
(even for an assumed spherically 
symmetric kinematic $\alpha$) the energy dependent 
quenching would result in a breaking of the spherical symmetry.  

In addition to this quenching, the $\alpha(r)$ profiles are 
perturbed  by
"blobs" of noise which are
considered constant within a correlation
time $\tau_{corr}$.
Physically, such a  noise term  can
be understood
as a consequence of changing boundary
conditions for the flow in the outer core, but also as a substitute
for the omitted influence of
higher multipole modes on the dominant axial dipole mode.

In summary, the $\alpha(r,\tau)$ profile  entering Eqs. (2) and (3)
is written as
\begin{eqnarray}
\alpha(r,\tau)&=&C \frac{1.916 \; (1-6 \; r^2+5 \; r^4)}{1+
E^{-1}_{mag,0} \left[ \frac{2 s_1^2(r,\tau)}{r^2}+
\frac{1}{r^2}\left( \frac{\partial (r s_1(r,\tau))}
{\partial r} \right)^2
+t_1^2(r,\tau) \right] }     \nonumber\\
&& +\xi_1(\tau) +\xi_2(\tau) \; r^2 +\xi_3(\tau) \; r^3+\xi_4(\tau) \; r^4 \; ,
\end{eqnarray}
where the noise correlation is given by
$\langle \xi_i(\tau) \xi_j(\tau+\tau_1)
\rangle = D^2 (1-|\tau_1|/\tau_{corr}) \Theta(1-|\tau_1|/\tau_{corr}) 
\delta_{ij}$.
$C$ is a normalized dynamo number,
$D$ is the noise strength, 
and $E_{mag,0}$ is a constant measuring the mean
magnetic field energy.

\begin{figure}[t]
\begin{center}
%   following commands fit figure into full page width..
\unitlength=\textwidth
\includegraphics[width=\textwidth]{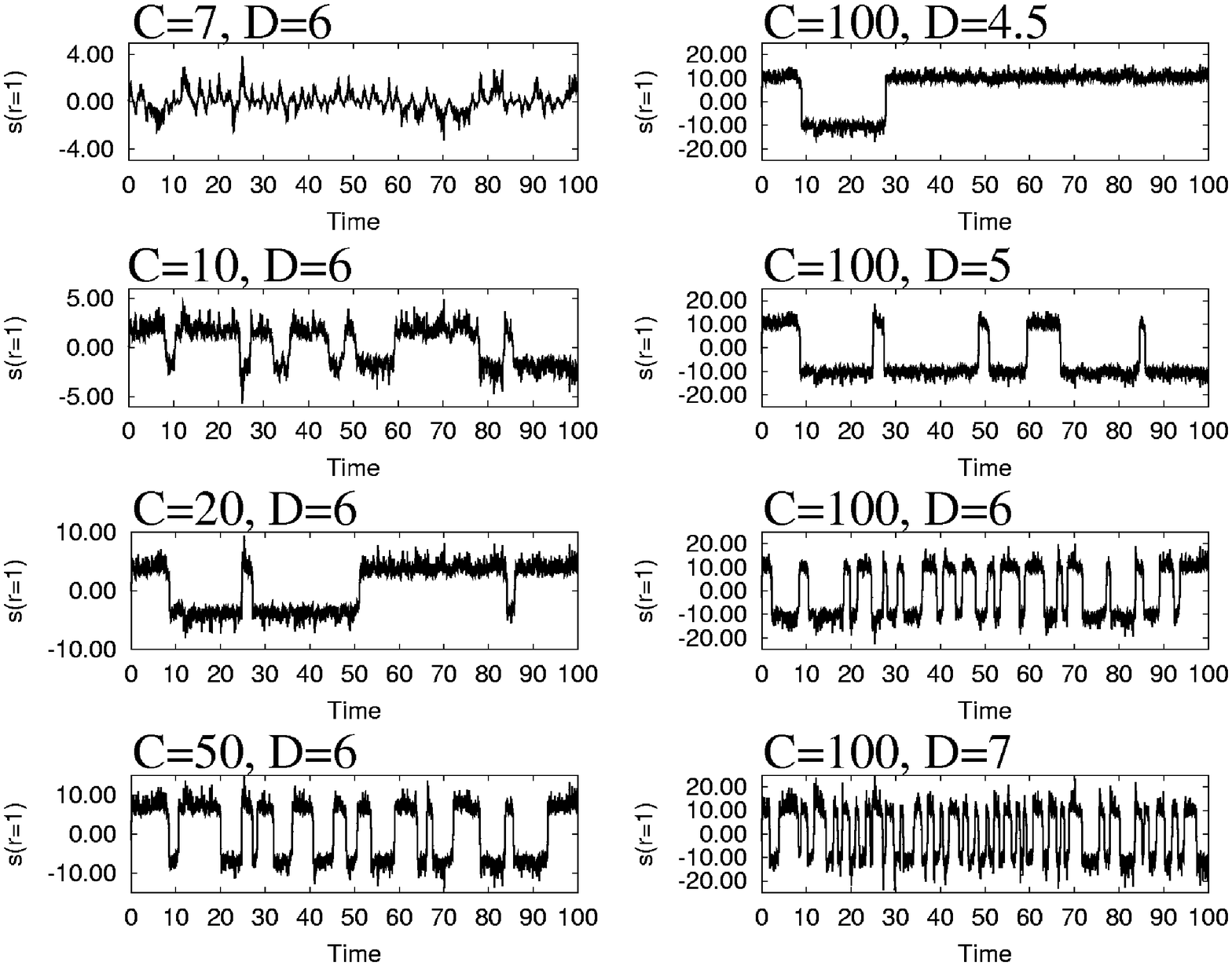}
 \caption{Time series for some selected values of $C$ and $D$.     }
\end{center}
\end{figure}

\section{Numerical results.}

We start with  the noise-free  case. 
Figure 1 shows
the magnetic field evolution according to the equation system
(2-4) for  $D=0$ and different
dynamo numbers $C$. The critical dynamo number is 6.78. 
The nearly harmonic oscillation for $C=6.8$ becomes more and more
saw-tooth shaped for increasing $C$, with a
pronounced asymmetry between the slow field decay and
the fast field recreation during the reversal.
At $C=7.24$ a transition to a steady dynamo has occurred.

In order to understand this behaviour, we examine
in Fig. 2 the evolution of the magnetic field 
within approximately half a period  of the 
anharmonic oscillation for the particular
value $C=7.237$ at which the saw-tooth shape is 
already very pronounced.
Figure 2a shows the time dependence of $s(r=1)$ during
this half period, with the typical slow decay 
and the fast recreation of the field.
This behaviour is analyzed in detail at 9 selected instants 
$\tau_i$ ($i=1...9$) for which 
the instantaneous fields $s(r,\tau_i)$ (Fig. 2d) and 
the corresponding
$\alpha(r,\tau_i)$ (Fig. 2e, 2f) are depicted.
The latter two plots reveal that $\alpha(r)$ undergoes
only slight changes during the oscillation and that
it comes very close to the unquenched, 
kinematic $\alpha_{kin}(r)$ (denoted by K)
when the magnetic field is small in the middle of the 
reversal.

It is instructive to plug the instantaneous 
$\alpha(r,\tau_i)$ profiles into an eigenvalue solver
(for which the time derivatives on the left hand
 sides of Eqs. (2) and (3)
are replaced by $\lambda s(r)$ and $\lambda t(r)$, respectively). 
Figure 2b and Fig. 2c show the resulting instantaneous 
growth rates  and frequencies  
during the half-oscillation.
Evidently, the reversal starts with a very slow 
field decay (slightly negative growth rate) 
which, however, accelerates itself
and drives the system into an oscillatory behaviour
(the frequency becomes different from zero for $2.0<\tau <4.5$).
Around $\tau=4$, when the
quenching of $\alpha$ is weak, the instantaneous growth rate 
reaches rather high values.
Later we will see that for strongly  supercritical dynamos 
these high growth rates result in a field dynamics
that is much faster than what would be expected 
from the diffusion time scale.

In Figure 3a we relate this magnetic field evolution to the existence of
exceptional points of the spectrum. It shows the
instantaneous growth rate which results from the
individual $\alpha(r,\tau_i)$ profiles, and
from the unquenched (kinematic) $\alpha(r)$ profile K.
In addition to the growth rates at the actual
dynamo number $C=7.237$ (indicated by the dashed vertical line), 
we have plotted
the whole growth rate curves in the interval between $4.5<C<7.5$.
At the exceptional point E$_k$ of the kinematic $\alpha_{kin}(r)$,  
the first eigenvalue branch K1 coalesces with the second 
branch K2 and both continue as
a pair of complex conjugate eigenvalues. For all 
other curves, only the exceptional
point is
indicated by $E_i$ whereas the branch of the second eigenvalue has been
omitted.
In this framework, a reversal can be described as follows:
At the instant 1, the growth rate  is located  close to the maximum of
the
non-oscillatory branch which is slightly below zero. The resulting
slow field decay accelerates itself, because the system moves down
(instant 2) from the
maximum of the real branch towards the exceptional point.
Then the system enters the oscillatory branch (3-7).
Finally the system moves back 
again
(8,9) but with opposite polarity.

The transition  point between oscillatory and steady dynamos ($C=7.239$) 
is characterized by the fact that the
maximum of the
non-oscillatory branch crosses the zero growth rate line. 
Beyond this point, the
field is growing rather than decaying, leading to
a stable fixed point somewhere to the left of the maximum of the
non-oscillatory branch, and hence to a non-oscillatory dynamo
(cf. the case $C=7.24$ in Fig. 1).
If noise comes into play it  will  soften
the sharp border between
oscillatory and steady dynamos. This means, in particular, that even 
above the transition point, the noise
can trigger a jump  to the right of the maximum from were
the described reversal process can start (Fig. 2).

\begin{figure}[t]
\begin{center}
%   following commands fit figure into full page width..
\unitlength=\textwidth
\includegraphics[width=\textwidth]{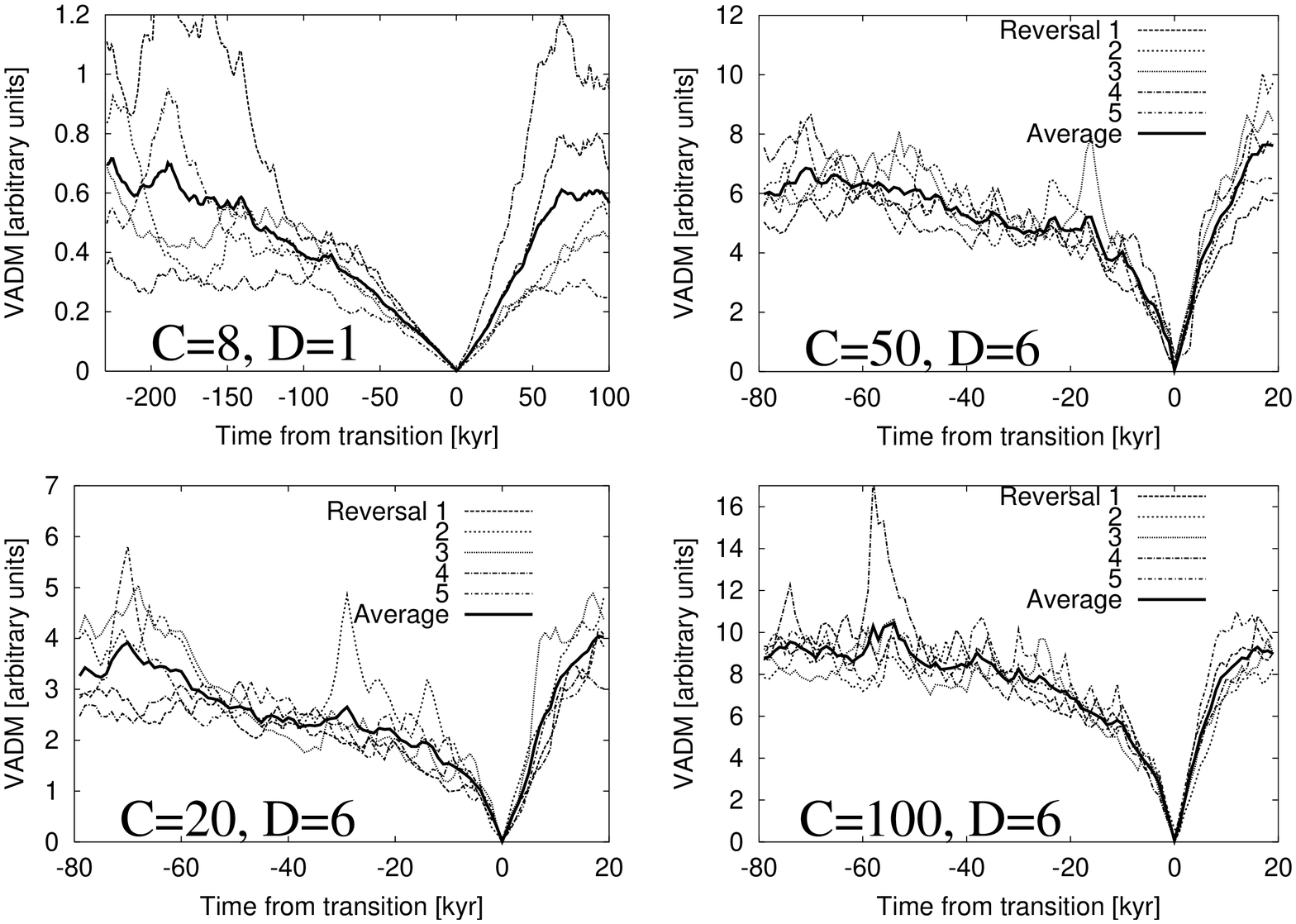}
\caption{Field intensity (virtual axial dipole moment) 
variations across five 
selected reversals, and their averages,   
computed for different values of $C$ and $D$. The time interval
includes 80 kyr before (negative values) and 20 kyr after
(positive values) a polarity transition. 
Note the quite different time scales for 
$C=8, D=1$ on one side, and for the  other three 
examples on the other side.}
\end{center}
\end{figure}

Having identified the exceptional point and the nearby 
local maximum as the spectral features that are responsible for both
the slow decay before and the fast recreation of the field  after
the polarity transition, one may  ask now 
why the operator of the actual geodynamo should possess
just such a special spectrum.

A preliminary answer to this question 
was given in \cite{EPSL}
and is  illustrated in Fig. 3b. 
Roughly speaking, highly supercritical dynamos
seem to have a  tendency to  saturate in a state
for which
the exceptional point and its associated local maximum
lie close to the zero growth rate line.
Interestingly, this happens independently on whether
the exceptional point in the original kinematic case
was above the
zero growth rate line 
or below it. Our example 
with $\alpha_{kin}(r)=1.916 \; C  \; (1-6 \; r^2+5 \; r^4)$ 
belongs to the second type. 
Starting from the kinematic $\alpha_{kin}(r)$,
for which the
exceptional point is well below the zero line, it
rises rapidly above zero to a maximum value, 
but for even higher $C$
it moves back in
direction to the
zero line.

Without noise  and for $C>7.239$, the position of the local maximum
above the zero growth rate line
leads to a steady dynamo. The presence of noise will
sometimes bring the actual growth rate below zero, 
and then the indicated 
reversal process can start.
A number of time series for different values of $C$ and $D$ 
is depicted in Fig. 4.

Figure 5 shows reversal details for four particular choices of
$C$ and $D$. For later comparison with paleomagnetic data
we have exhibited five typical reversals, and their averages,
from 80 kyr before  the polarity transition until 20 kyr after
(except the case $C=8, D=1$ for which the reversal takes much longer). 

\section{Comparison with paleomagnetic data.}

In this section we will validate if our model can 
be fitted to real paleomagnetic data. For this purpose we 
have used 
recently published material 
about five  reversals which occurred during the last two million years
\cite{VALET2005}. Actually, the
data shown in Fig. 6a have been 
extracted from 
Fig. 4 of \cite{VALET2005}.
In all five curves, as well as in their averages, the  asymmetry 
of the reversal process is clearly visible. The dominant 
features are a field decay over a period  of 
50-80 kyr and a  rather sharp field recreation within 5-10 kyr.
It has been an  old-standing puzzle to explain in particular this
fast recreation when a diffusive timescale of 200 kyr has to 
be taken into
account. A possible solution of this problem is to assume
a turbulent resistivity which is much larger than the
molecular resistivity (e.g., by a factor 15-20 in \cite{HOYNG}).

The comparison of the real data with the numerically time series 
in Fig. 6b shows that this assumption 
of turbulent resistivity is by no means 
necessary. 
Apart from the slightly supercritical 
case $C=8, D=1$ which
exhibits a much to slow magnetic field evolution, the other
examples with $C=$20, 50, 100 and  $D=6$ show very realistic time
series with the typical slow decay and  fast recreation.
As noted above, the fast recreation results from the fact 
that in a small intervall during the transition the dynamo 
operates with an nearly
unquenched $\alpha(r)$ profiles which yields, in case that 
the dynamo is strongly supercritical,
rather high growth rates.
\begin{figure}[t]
\begin{center}
%   following commands fit figure into full page width..
\unitlength=\textwidth
\includegraphics[width=\textwidth]{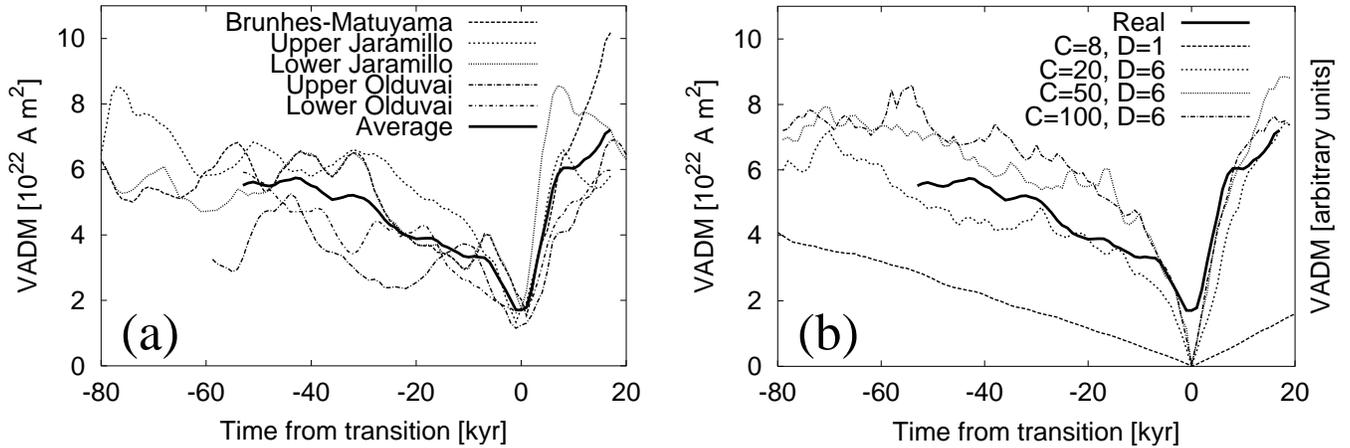}
 \caption{Comparison of paleomagnetic reversal data and numerically
simulated ones. (a) Virtual axial dipole moment (VADM) during
the 80 kyr preceeding  and the 20 kyr following a polarity transition
for five reversals from the last 2 million years 
(data extracted from \cite{VALET2005}), 
and their average. (b) Comparison of the average curve of (a) with the 
four average curves of Fig. 5.  The field scale for the numerical
values has been fixed in such a way that the intensity {\it in the 
non-reversing} periods matches approximately the observed values.}
\end{center}
\end{figure}

\section{Conclusions}

Based on former results in \cite{PRL} and \cite{EPSL},
we have shown that a simple but strongly  supercritical 
$\alpha^2$ dynamo model exhibits a number of features which 
are typical for Earth's magnetic field reversals, in particular 
an asymmetric shape  and correct timescales for the field 
decay and field recreation.

As it does not include the necessary  North-South asymmetry of 
$\alpha$ we cannot claim that our model is an appropriate model of the
geodynamo.
However, recent papers by Giesecke et al. have shown that
a) even in such realistic models $\alpha$ may exhibit a sign change
along the radius \cite{GIES} and b) that such models
can also give rise to reversals \cite{GIESAN}.
Therefore, it seems worthwhile to identify the indicated reversal
scenario in this and in more complicated geodynamo models.

%-----------------------------------------------------------------------

\end{document}